\documentclass[prl,twocolumn,showpacs,superscriptaddress,preprintnumbers,floatfix]{revtex4}

\usepackage{ifpdf}
\usepackage{hyperref}
\usepackage{dcolumn}
\usepackage{url}
\usepackage{amsmath}
\usepackage{amscd}
\usepackage{amsfonts}
\usepackage{amssymb}
\usepackage{bm}   
\usepackage{bbm}   
\usepackage{verbatim}
\usepackage{stmaryrd}
\usepackage{amsthm}
\usepackage{color}

\ifpdf
  \usepackage[pdftex]{graphicx}         
\else
  \usepackage[dvips]{graphicx}
\fi

\theoremstyle{plain}   
\theoremstyle{plain} 	
\theoremstyle{plain} 	
\theoremstyle{plain} 	
\theoremstyle{plain} 	
\theoremstyle{plain}	
\theoremstyle{plain}	 
\theoremstyle{plain}	

\newcommand{\figref}[1]{Fig.~\ref{#1}}

\newcommand{\refcite}[1]{Ref.~\cite{#1}}

\newcommand{\cryptic}[1]{$#1$\protect\nobreakdash-cryptic}

\newcommand{\eM}     {$\epsilon$\protect\nobreakdash-machine}
\newcommand{\eMs}    {$\epsilon$\protect\nobreakdash-machines}

\newcommand{\MeasAlphabet}	{\mathcal{A}}
\newcommand{\MeasSymbol}   { {X} }
\newcommand{\meassymbol}   { {x} }

\newcommand{\Future}	{ \overrightarrow{\MeasSymbol} }

\newcommand{\CausalState}	{ \mathcal{S} }

\newcommand{\causalstate}	{ \sigma }
\newcommand{\causalstateprime}	{ \sigma^{\prime} }

\newcommand{\Cmu}		{C_\mu}

\newcommand{\EE}		{{\bf E}}

\newcommand{\PC}		{\chi}

\newcommand{\forward}{+}
\newcommand{\reverse}{-}

\newcommand{\FutureCausalState}	{ {\CausalState}^{\forward} }

\newcommand{\PastCausalState}	{ {\CausalState}^{\reverse} }

\newcommand{\eMachine}	{ M }
\newcommand{\FutureEM}	{ {\eMachine}^{\forward} }

\newcommand{\FutureEps}	{ \epsilon^{\forward} }

\newcommand{\TR}{\mathcal{T}}
\newcommand{\MSP}{\mathcal{U}}

\begin{document}

\title{Information Accessibility and Cryptic Processes:\\
Linear Combinations of Causal States}

\author{John R. Mahoney}
\email{jrmahoney@ucdavis.edu}
\affiliation{Complexity Sciences Center and Physics Department,
University of California at Davis, One Shields Avenue, Davis, CA 95616}

\author{Christopher J. Ellison}
\email{cellison@cse.ucdavis.edu}
\affiliation{Complexity Sciences Center and Physics Department,
University of California at Davis, One Shields Avenue, Davis, CA 95616}

\author{James P. Crutchfield}
\email{chaos@cse.ucdavis.edu}
\affiliation{Complexity Sciences Center and Physics Department,
University of California at Davis, One Shields Avenue, Davis, CA 95616}
\affiliation{Santa Fe Institute, 1399 Hyde Park Road, Santa Fe, NM 87501}

\date{\today}

\bibliographystyle{unsrt}

\begin{abstract}
We show in detail how to determine the time-reversed representation of a
stationary hidden stochastic process from linear combinations of its
forward-time \eM\ causal states. This also gives a check for the \cryptic{k}
expansion recently introduced to explore the temporal range over which
internal state information is spread.
\end{abstract}

\preprint{Santa Fe Institute Working Paper 09-06-XXX}
\preprint{arxiv.org:0906.XXXX [physics.cond-mat]}

\maketitle


\section{Introduction}

We introduced a new system ``invariant''---the \emph{crypticity} $\PC$---for
stationary hidden stochastic processes to capture how much internal state
information is directly accessible (or not) from observations
\cite{Crut08a,Crut08b,Maho09a}. Two approaches to calculate $\PC$ were given.
The first, reported in \refcite{Crut08a} and \refcite{Crut08b}, used the
so-called \emph{mixed-state} method, which employs linear combinations of a
process's forward-time \eM. The second, appearing in \refcite{Maho09a},
developed a systematic expansion $\PC(k)$ as a function of the length $k$ of
observed sequences over which internal state information can be extracted.
The mixed-state method is the most efficient way to calculate crypticity and
other important system properties, such as the excess entropy $\EE$, since it
avoids having to write out all of the terms required for calculating $\PC(k)$.
It also does not rely on knowing in advance a process's cryptic order.

As such, we reported results in \refcite{Maho09a} that use the mixed-state
method to, in a sense, calibrate the $\PC(k)$ expansion and to understand its
convergence.

Here we provide the calculational details behind those results. Generally,
though, the goal is to
find out what a stochastic process looks like when scanned in the ``opposite''
time direction. Specifically, starting with a given \eM\ $M$ of a process,
calculate its reverse-time representation $M^-$. (The latter is not always
minimal and so not, in that case, an \eM.) This is done in two steps: (i)
time-reverse $M$, producing $\widehat{M} = \mathcal{T}(M)$, and (ii) convert
$\widehat{M}$ to a unifilar presentation $\mathcal{U}(\widehat{M})$ using mixed
states, which are linear combinations of the states of $\widehat{M}$.

\begin{figure}[th]
\centering
\includegraphics{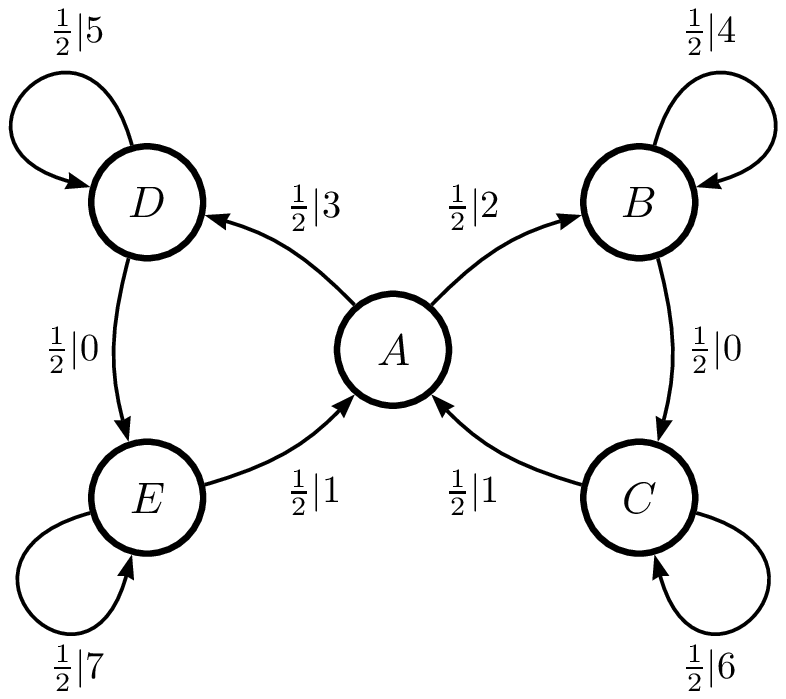}
\caption{A \cryptic{2} process: The \eM\ representation of the Butterfly
  Process. Edge labels $t|\meassymbol$ give the probability
  $t = T^{(\meassymbol)}_{\causalstate\causalstateprime}$ of making a
  transition and from causal state $\causalstate$ to causal state
  $\causalstateprime$ and seeing symbol $\meassymbol$.
  }
\label{fig:ButterflyProcess}
\end{figure}

In the following, we show how to implement these steps for the various example
processes presented in \refcite{Maho09a}: the Butterfly, Restricted Golden Mean,
and Nemo Processes. We jump directly into the calculations, assuming the reader
is familiar with Refs. \cite{Crut08a}, \cite{Crut08b}, and \cite{Maho09a}.
Those references provide, in addition, more discussion and motivation and
reasonable list of citations.

\section{Butterfly Process}

Figure \ref{fig:ButterflyProcess} shows the \eM\ for \refcite{Maho09a}'s
Butterfly process---an output process over eight symbols
$\MeasAlphabet = \{0,1,\ldots,7\}$.

\begin{figure}[th]
\centering
\includegraphics{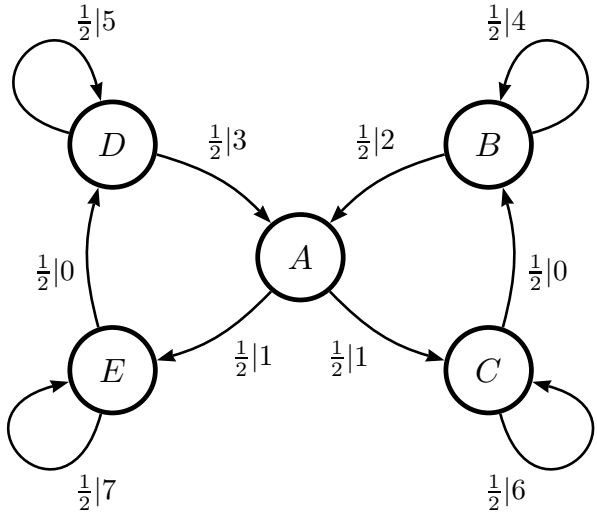}
\caption{Time-reversed Butterfly Process.}
\label{fig:TRButterflyProcess}
\end{figure}

Since its transition matrices are doubly stochastic, the stationary state
distribution is uniform. This immediately gives its stored information: the
statistical complexity is $\Cmu = \log_2(5)$ bits. It also makes the
construction of the time-reverse machine straightforward: We simply reverse
the directions of all the arrows. (See Fig. \ref{fig:TRButterflyProcess}.)
Note that the time-reverse presentation is no longer unifilar and, therefore,
it is not the reversed process's \eM. 

Due to this we must calculate the mixed-state presentation to find a unifilar
presentation. The calculated mixed states and the words which induce them are
given in Table \ref{tab:ButterflyProcessMixedStates}.

\begin{table}[th]
\begin{center}
\begin{tabular}{|c|c|}
\hline
Allowed Words & $\mu$ or Previous Word\\
\hline
0 & (0,$\frac{1}{2}$,0,$\frac{1}{2}$,0)\\
1 & (0,0,$\frac{1}{2}$,0,$\frac{1}{2}$)\\
2 & (1,0,0,0,0)\\
3 & 2\\
4 & (0,1,0,0,0)\\
5 & (0,0,0,1,0)\\
6 & (0,0,1,0,0)\\
7 & (0,0,0,0,1)\\
02 & 2\\
03 & 2\\
04 & 4\\
05 & 5\\
10 & 0\\
16 & 6\\
17 & 7\\
21 & 1\\
42 & 2\\
44 & 4\\
53 & 2\\
55 & 5\\
60 & 4\\
66 & 6\\
70 & 5\\
77 & 7\\
\hline
\end{tabular}
\end{center}
\caption{Calculating the time-reversed Butterfly Process's \eM\ via the forward
  \eM's mixed states. The $5$-vector denotes the mixed-state distribution
  $\mu(w)$ reached after having seen the corresponding allowed word $w$. If
  the word leads to a unique state with probability one, we give instead the
  state's name.
  }
\label{tab:ButterflyProcessMixedStates}
\end{table}

\begin{figure}[th]
\centering
\includegraphics{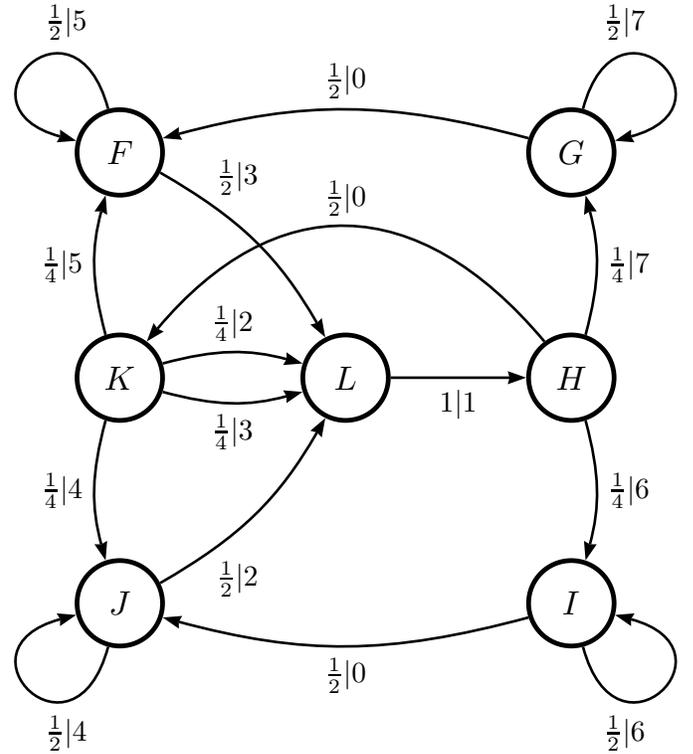}
\caption{Reverse Butterfly Process.
  }
\label{fig:ReverseButterflyProcess}
\end{figure}

The result is the reverse \eM\ shown in Fig.~\ref{fig:ReverseButterflyProcess}.
Note that it has two more states than the original (forward) \eM\ of
Fig.~\ref{fig:ButterflyProcess}.

The stationary distribution of this reversed machine is
$\pi = (0.1, 0.2, 0.2, 0.15, 0.15, 0.1, 0.1)$. Now we are in position
to calculate $\EE$ using the result of \refcite{Crut08a}:
\begin{align}
\EE &= \Cmu - \PC \\
\EE &= \Cmu - H[\FutureCausalState | \Future] \\
    &= \Cmu - H[\FutureCausalState | \PastCausalState =
	\FutureEps(\Future)] ~.
\end{align}
In this case, we find a crypticity of:
\begin{align*}
  \PC & = H[\FutureCausalState | \PastCausalState] \\
    & =  0.1 H[(0,\frac{1}{2},0,\frac{1}{2},0)]
		+ 0.2 H[(0,0,\frac{1}{2},0,\frac{1}{2})] \\
	& \quad + 0.2 H[(1,0,0,0,0)]
		+ 0.15 H[(0,1,0,0,0)] \\
	& \quad + 0.15 H[(0,0,0,1,0)]
		+ 0.1 H[(0,0,1,0,0)] \\
	& \quad + 0.1 H[(0,0,0,0,1)] \\
    & = 0.1 + 0.2 \\
	& = 0.3 ~\mathrm{bits}.
\end{align*}
So, $\EE = \log_2 (5) - 0.3 \approx 2.0219$ bits, in accord with the result
calculated via Thm. 1 of \refcite{Maho09a}.

\section{Restricted Golden Mean Process}

\begin{figure}[th]
\begin{center}
\includegraphics{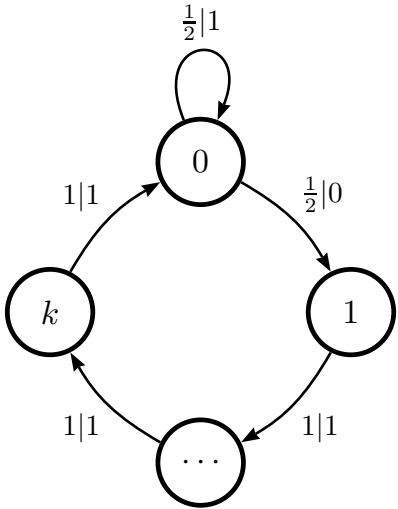}
\caption{The \eM\ for the Restricted Golden Mean Process.}
\label{fig:RestrictedGM}
\end{center}
\end{figure}

For reference, we give the family of labeled transition matrices
for the binary Restricted Golden Mean Process (RGMP):
\begin{align*}
T^{(0)} & = \begin{pmatrix}
0 & \frac{1}{2} & 0 & 0 & 0 &\cdots\\
0 & 0 & 0 & 0 & 0 & \cdots\\
0 & 0 & 0 & 0 & 0 & \cdots\\
0 & 0 & 0 & 0 & 0 & \cdots\\
\vdots & \vdots & \vdots & \vdots & \vdots & \ddots\\
0 & 0 & 0 & 0 & 0 & \cdots
\end{pmatrix}
\end{align*}
and
\begin{align*}
T^{(1)} & = \begin{pmatrix}
\frac{1}{2} & 0 & 0 & 0 & 0 &\cdots\\
0 & 0 & 1 & 0 & 0 & \cdots\\
0 & 0 & 0 & 1 & 0 & \cdots\\
0 & 0 & 0 & 0 & 1 & \cdots\\
\vdots & \vdots & \vdots & \vdots & \vdots & \ddots\\
1 & 0 & 0 & 0 & 0 & \cdots
\end{pmatrix} ~.
\end{align*}
Its \eM\ is given in Fig. \ref{fig:RestrictedGM} and its stationary
distribution is:
\begin{align*}
\pi = \left(\frac{2}{k+2}, \frac{1}{k+2}, \frac{1}{k+2},
  \ldots, \frac{1}{k+2}\right) ~.
\end{align*}

\begin{figure}[th]
\begin{center}
\includegraphics{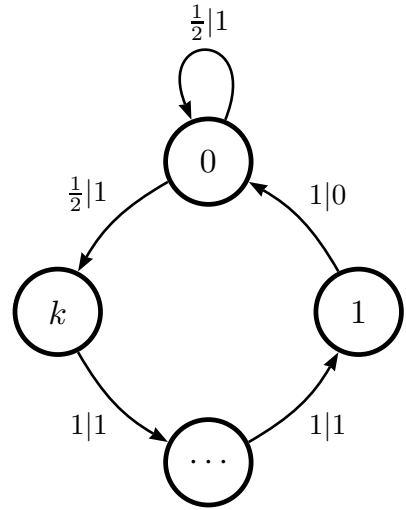}
\caption{Time-reversed presentation of the Restricted Golden Mean Process.}
\label{fig:TRRestrictedGM}
\end{center}
\end{figure}

\begin{figure}[th]
\begin{center}
\includegraphics{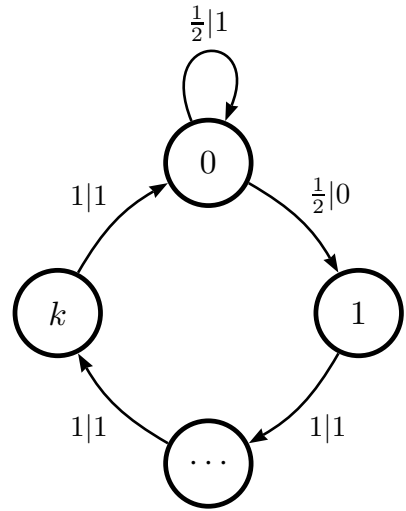}
\caption{Reverse Restricted Golden Mean Process.}
\label{fig:ReverseRestrictedGM}
\end{center}
\end{figure}

\begin{table}
\begin{center}
\begin{tabular}{|c|c|}
\hline
Allowed Words & $\mu$ or Previous Word\\
\hline
0 & $(1, 0^k)$\\
1 & $(\frac{1}{k+1}, \frac{1}{k+1}, \ldots, \frac{1}{k+1})$\\
01 & $(\frac{1}{2}, 0^{k-1}, \frac{1}{2})$\\
10 & 0\\
11 & $\frac{1}{k}(\frac{1}{2},1,1,\ldots,1,\frac{1}{2})$\\
\vdots & \vdots \\
0$(1)^n$ for $1 \leq n \leq k$ & $(\frac{1}{2^{n}}, 0^{k-n}, \frac{1}{2^1} \frac{1}{2^2} \frac{1}{2^3}, \ldots,  \frac{1}{2^{n}})$\\
1$(1)^n$ for $1 \leq n \leq k$ & $\frac{1}{k-n+1}(\frac{1}{2^{n}}, 1^{k-n}, \frac{1}{2^1} \frac{1}{2^2} \frac{1}{2^3}, \ldots,  \frac{1}{2^{n}})$\\
0$(1)^{k}$ & $(\frac{1}{2^{k}}, \frac{1}{2^1} \frac{1}{2^2} \frac{1}{2^3}, \ldots,  \frac{1}{2^{k}})$\\
1$(1)^{k}$ & 0$(1)^{k}$\\
0$(1)^{k}$0 & 0\\
0$(1)^{k}$1 & 0$(1)^{k}$\\
\hline
\end{tabular}
\end{center}
\caption{Calculating the reversed RGMP using mixed states over the
  \eM\ states.
  }
\label{tab:RestrictedGMMixedStates}
\end{table}

Through other methods, we can show that the RGMP is reversible. We ``push'' RGMP
to an edge machine presentation and ``pull'' $\TR$(RGMP) also the same type of
presentation. (An edge machine presentation of a machine $M$ has states that
are the edges of $M$.) These machines are the same. Therefore, the forward and
reverse \eMs\ are the same and, moreover, we can use the same mixed-state
inducing word list. It is easy to see that one such list is
$(0,01,011,\ldots,01^k)$. Table \ref{tab:RestrictedGMMixedStates} gives the
mixed states for these allowed words. It is also reasonably clear from the above
mixed-state presentation that these correspond to the recurrent causal states
for the time-reversed process's \eM.

With this, we can now compute $\PC$ using
$H[\FutureCausalState | \PastCausalState]$, as follows:
\begin{align*}
H[\FutureCausalState | \PastCausalState = 0] &=  H[(1, 0^k)] = 0 ~\mathrm{and}\\
H[\FutureCausalState | \PastCausalState = 0(1)^n]
  &= H[(\frac{1}{2^{n}}, 0^{k-n}, \frac{1}{2^1} \frac{1}{2^2}
  \frac{1}{2^3}, \ldots,  \frac{1}{2^{n}})] ~.
\end{align*}
So that, in general, we have:
\begin{align*}
H[\FutureCausalState | \PastCausalState]
  & = \sum_{n=1}^{k-1}{\frac{1}{k+2} H[(\frac{1}{2^{n}}, 0^{k-n},
  \frac{1}{2^1} \frac{1}{2^2} \frac{1}{2^3}, \ldots,  \frac{1}{2^{n}})]}\\
  & \quad
  + \frac{2}{2+k} H[(\frac{1}{2^{k}}, \frac{1}{2^1} \frac{1}{2^2} \frac{1}{2^3},
  \ldots,  \frac{1}{2^{k}})] ~.
\end{align*}
It can then be shown that:
\begin{align*}
H[(\frac{1}{2^{n}} & , 0^{k-n}, \frac{1}{2^1} \frac{1}{2^2} \frac{1}{2^3},
  \ldots, \frac{1}{2^{n}})] \\
  & = H[(\frac{1}{2^{n}}, \frac{1}{2^1} \frac{1}{2^2} \frac{1}{2^3}, \ldots,
  \frac{1}{2^{n}})] \\
  & = 2 - 2^{(1-n)} ~.
\end{align*}
Therefore, returning to the causal-state-conditional entropy of interest,
we have:
\begin{align*}
H[\FutureCausalState | \PastCausalState] &=
  \frac{1}{k+2}\sum_{n=1}^{k-1}{(2 - 2^{(1-n)})} + \frac{2}{2+k} (2 - 2^{(1-k)})\\
  &= \frac{1}{k+2}(2(k-1) + 2(2-2^{1-k}) - (2-2^{2-k}))\\
  &= \frac{2k}{k+2} ~.
\end{align*}
With a few more steps, we arrive at our destination---the
RGMP's informational quantities:
\begin{align*}
\Cmu & = \log2(k+2) - \frac{2}{k+2} ~,\\ 
\PC  & = \frac{2k}{k+2}, ~\mathrm{and}\\
\EE  & = \log2(k+2) - \frac{2(k+1)}{k+2} ~.
\end{align*}

\begin{figure}[th]
\begin{center}
\includegraphics{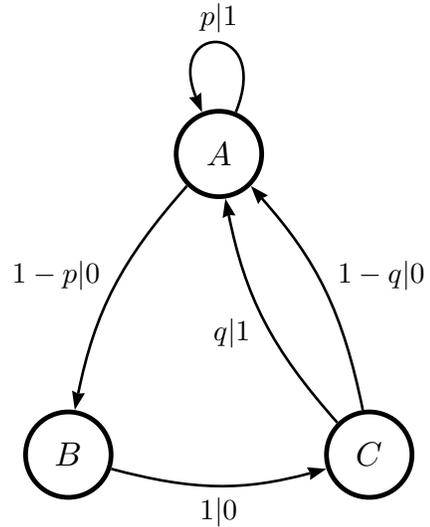}
\caption{The \eM\ for the $\infty$-cryptic Nemo Process.}
\label{fig:Nemo}
\end{center}
\end{figure}

\section{Nemo Process}

We now demonstrate how to calculate $\PC$ and $\EE$ for \refcite{Maho09a}'s
$\infty$-cryptic process---the Nemo Process---using mixed-state methods. As
emphasized in \refcite{Maho09a}, the \cryptic{k} expansion there cannot be
applied in this case. Thus, the Nemo Process demonstrates that
Refs. \cite{Crut08a} and \cite{Crut08b}'s mixed-state method is essential.

Figure \ref{fig:Nemo} shows $\FutureEM$, the \eM\ for the forward-scanned
Nemo Process. Its transition matrices are:
\begin{align*}
T^{(0)} &= 
\bordermatrix{%
  & A & B & C \cr
A & 0 & 1-p & 0 \cr
B & 0 & 0 & 1 \cr
C & 1-q & 0 & 0
} \textrm{ and }\\
T^{(1)} &=
\bordermatrix{%
  & A & B & C \cr
A & p & 0 & 0 \cr
B & 0 & 0 & 0 \cr
C & q & 0 & 0
}.
\end{align*}
The stationary state distribution is the normalized left-eigenvector of 
$T \equiv T^{(0)} + T^{(1)}$ and is given by:
\begin{align*}
\Pr(\FutureCausalState) \equiv \pi^+ = \frac{1}{3-2p}
\bordermatrix{%
 & A & B & C \cr
 & 1 & 1-p & 1-p
}.
\end{align*}
Then, the statistical complexity is the Shannon entropy over these states:
\begin{align*}
\Cmu &= H[\FutureCausalState]  \\
           & = \log_2(3-2p) - \frac{2(1-p)}{3-2p} \log_2 (1-p) ~.
\end{align*}

\begin{figure}[th]
\begin{center}
\includegraphics{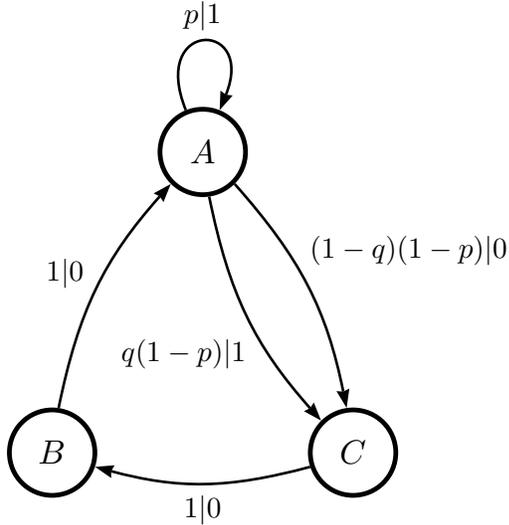}
\caption{The time-reversed presentation, $\widetilde{M}^+ = \TR(\FutureEM)$,
  of the Nemo Process.
  }
\label{fig:TRNemo}
\end{center}
\end{figure}

The next step is to construct the 
time-reversed presentation $\widetilde{M}^+ = \TR(\FutureEM)$, shown in 
\figref{fig:TRNemo}.  The transition matrices of this machine are:
\begin{align*}
\widetilde{T}^{(0)} &=
    \bordermatrix{
    & A & B & C \cr
  A & 0 & 0 & (1-q)(1-p)\cr
  B & 1 & 0 & 0\cr
  C & 0 & 1 & 0\cr
} \textrm{ and} \\
\widetilde{T}^{(1)} &=
    \bordermatrix{
    & A & B & C \cr
  A & p & 0 & q(1-p)\cr
  B & 0 & 0 & 0\cr
  C & 0 & 0 & 0\cr
}.
\end{align*}

\begin{figure}
\centering
\includegraphics{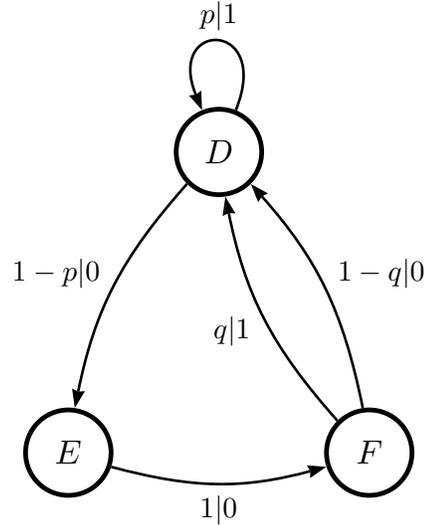}
\caption{The reverse \eM\ for the Nemo Process.
  }
\label{fig:ReverseNemo}
\end{figure}

Finally, we construct the mixed-state presentation of the time-reversed
presentation, $\MSP(\widetilde{M}^+)$, which is shown in 
\figref{fig:ReverseNemo}. On doing so, we obtain the following
mixed states:
\begin{align*}
D \equiv \nu(1) &= \frac{1}{p + q - pq} 
\bordermatrix{
 & A & B & C \cr
 & p & 0 & q(1-p)
} ~,\\
E \equiv \nu(01) &= \frac{1}{p + q - pq} 
\bordermatrix{
 & A & B & C \cr
 & 0 & q & p(1-q)
} ~, ~\mathrm{and}\\
F \equiv \nu(001) &= \frac{1}{p + q - pq} 
\bordermatrix{
 & A & B & C \cr
 & q & p(1-q) & 0 
} ~.
\end{align*}

These mixed states form the reverse \eM\ causal states, which are exactly the
same as the forward \eM. Thus, the Nemo Process is causally reversible. The
mixed states are distributions giving the probabilities of the forward causal
states conditioned on a reverse causal state:
\begin{align*}
\Pr(\FutureCausalState|\PastCausalState) &= \frac{1}{p + q - pq}
\bordermatrix{
  & A & B & C \cr
D & p & 0 & q(1-p) \cr
E & 0 & q & p(1-q) \cr
F & q & p(1-q) & 0 \cr
} ~.
\end{align*}
We use this to directly compute:
\begin{align*}
H[\FutureCausalState|\PastCausalState] 
    &= \frac{1}{3-2p} \biggl[ \frac{p}{p+q-pq} \log_2 \left(\frac{p+q-pq}{p}\right) \biggl.\\
    &\quad \biggl. + \frac{q(1-p)}{p + q - pq} \log_2 \left(\frac{p+q-pq}{q(1-p)}\right) \biggr]\\
    &\quad +\frac{2(1-p)}{3-2p} \biggl[ \frac{q}{p+q-pq} \log_2 \left(\frac{p+q-pq}{q}\right) \biggr.\\
    &\quad \biggl. + \frac{p(1-q)}{p+q-pq} \log_2
	\left(\frac{p+q-pq}{p(1-q)}\right) \biggr] ~.
\end{align*}
Finally, we have:
\begin{align*}
\EE &= \Cmu - H[\FutureCausalState|\PastCausalState]\\
    &= \log_2(3-2p) - \frac{2(1-p)}{3-2p} \log_2 (1-p) \\
    &\quad - \frac{1}{3-2p} \biggl[ \frac{p}{p+q-pq} \log_2 \left(\frac{p+q-pq}{p}\right) \biggl.\\
    &\quad \biggl. + \frac{q(1-p)}{p + q - pq} \log_2 \left(\frac{p+q-pq}{q(1-p)}\right) \biggr]\\
    &\quad +\frac{2(1-p)}{3-2p} \biggl[ \frac{q}{p+q-pq} \log_2 \left(\frac{p+q-pq}{q}\right) \biggr.\\
    &\quad \biggl. + \frac{p(1-q)}{p+q-pq} \log_2
	\left(\frac{p+q-pq}{p(1-q)}\right) \biggr] ~.
\end{align*}

\section{Conclusion}

The detailed calculations make evident that Refs. \cite{Crut08a} and
\cite{Crut08b}'s mixed-state method gives a new level of direct analysis for
the informational properties of stationary stochastic processes, such as the
crypticity and the excess entropy. The complementary approach given by the
crypticity expansion $\PC(k)$ is useful in understanding information
accessibility---how internal state information is spread over time in
measurement sequences \cite{Maho09a}. Nonetheless, while $\PC(k)$ can be
calculated in particular finite cases, the mixed-state method is the most
general and efficient method.

\bibliography{ref,chaos}

\begin{thebibliography}{1}

\bibitem{Crut08a}
J.~P. Crutchfield, C.~J. Ellison, and J.~Mahoney.
\newblock Time's barbed arrow: {Irreversibility}, crypticity, and stored
  information.
\newblock {\em submitted}, 2009.
\newblock arxiv.org:0902.1209 [cond-mat].

\bibitem{Crut08b}
C.~J. Ellison, J.~R. Mahoney, and J.~P. Crutchfield.
\newblock Prediction, retrodiction, and the amount of information stored in the
  present.
\newblock {\em arxiv: 0905.3587 [cond-mat]}, 2009.

\bibitem{Maho09a}
J.~R. Mahoney, C.~J. Ellison, , and J.~P. Crutchfield.
\newblock Information accessibility and cryptic processes.
\newblock {\em submitted}, 2009.
\newblock arxiv.org:0905.4787 [cond-mat].

\end{thebibliography}

\end{document}